\documentclass[conference]{IEEEtran}
\IEEEoverridecommandlockouts
\usepackage{csquotes}
\usepackage{cite}
\usepackage{amsmath,amssymb,amsfonts}
\usepackage{algorithmic}
\usepackage{graphicx}
\usepackage{textcomp}
\usepackage{xcolor}
\usepackage{paralist}
\usepackage{subcaption}
\usepackage{xurl}
\usepackage{hyperref}
\hypersetup{hidelinks}

\usepackage[english]{babel}
\usepackage[utf8x]{inputenc}
\usepackage[T1]{fontenc}

\usepackage[a4paper,top=3cm,bottom=2cm,left=2.5cm,right=2.5cm,marginparwidth=1.75cm]{geometry}
\usepackage{multicol}
\usepackage{parskip}

\usepackage{amssymb}
\usepackage{amsmath}
\usepackage{tikz}
\usetikzlibrary{positioning}
\usetikzlibrary{arrows.meta}
\usetikzlibrary{shapes.geometric}
\colorlet{cheatlag}{orange!80!black}
\colorlet{cheatdos}{purple!80!black}

\tikzset{
    cheatlag/.style={
        draw=cheatlag,
        fill=cheatlag!30,
        regular polygon,
        regular polygon sides=3,
        minimum size=4mm
    },
    cheatdos/.style={
        draw=cheatdos,
        thick,
        scale=2
    },
    movement/.style={
        thick,
        pathcolor,
        -{Latex[length=3mm, width=2.5mm]}
    }
}

\usepackage{tabularx}

\usepackage{longtable}
\usepackage{array}
\usepackage{booktabs}

\usepackage{listings}
\usepackage{inconsolata} % optional, nicer monospace font

\usepackage{listings}
\usepackage{xcolor}
\usepackage{graphicx}
\usepackage{float}
\usepackage{svg}

\definecolor{lightgray}{gray}{0.95}

\title{Cheating in Multiplayer Online Games: a Dataset} 

\author{\IEEEauthorblockN{Hugo Bertin}
\IEEEauthorblockA{\textit{Univ Rennes, CNRS, INRIA, IRISA} \\
Rennes, France \\
hugo.bertin@irisa.fr}
\and
\IEEEauthorblockN{Marc Dacier}
\IEEEauthorblockA{\textit{KAUST} \\
Thuwal, Saudi Arabia \\
marc.dacier@kaust.edu.sa}
\and
\IEEEauthorblockN{Yérom-David Bromberg}
\IEEEauthorblockA{\textit{Univ Rennes, CNRS, INRIA, IRISA} \\
Rennes, France \\
david.bromberg@irisa.fr}
}

\begin{document}

\maketitle

\thispagestyle{plain}
\pagestyle{plain}

%% Abstract
\begin{abstract}
Cheating poses a significant threat to the Multiplayer Online Games (MOG) industry by degrading player satisfaction and undermining the fairness in competitive gaming. Despite efforts to develop mitigation techniques, cheating remains difficult to detect and prevent in practice. In particular, a class of cheats based on network flow disruption remains unsolvable. To find out how to detect such attacks we need access to representative labelled data. However, no such dataset exists. 

To address this gap, we leverage an experimental framework that combines a multiplayer online game with a plug-in capable of both reproducing cheating attacks and collecting logs at two levels: network and application-layer. 

This paper presents a dataset compiling records of game sessions played by both real players and automated game clients, with cheating actions explicitly logged. To the best of our knowledge, this is the first dataset that provides logs of network flow disruption cheats. While it includes such network-based cheats, it is not limited to them and also contains records of more commonly studied cheats, such as aimbots and wallhacks. This dataset can be used by researchers in academia and industry seeking to develop cheating detection mechanisms for online games. Furthermore, it is designed to be evolutive and can be enriched by others creating their own data traces with the proposed framework.

\end{abstract}
    
\begin{table*}[ht]
    \centering
    \begin{tabular}{l p{8.5cm}}
        \hline
        Subject                  & \textbf{Computer Sciences}           \\
        Specific subject area    & Cybersecurity. Cheating Detection in Multiplayer Online Games.                                     \\
        Type of data             & Raw Network Traffic Traces (pcap files), Filtered Network Traffic Traces (Protocol Buffers-encoded binary format), and Filtered application-level game logs (Protocol Buffers-encoded binary format).   \\
        Data collection          & Data collection relied on an experimental framework that combines a multiplayer online game with a plug-in to enable log collection and controlled cheat execution. The game was distributed to players in the form of a game client executable. The game server was running at a cloud provider infrastructure.
        
        This setup enabled an experimental data collection campaign organized in game sessions, including participants from geographically distributed locations. The data was collected at both the game server and the clients.

        Additional logs were collected using automated game clients to reproduce pre-defined scenarios of gameplay and cheating multiple times. 
                          \\
        Data source location     &  \textit{Server-side data}: collected on the game server running on a virtual private server hosted in Gravelines, France by OVH provider.

        \textit{Client-side data}:
        \begin{itemize}
            \item Human-played sessions: data were collected on game client computers owned by participants with distributed geographic locations, as specified in the dataset. 
            \item Scripted sessions:  Client data were collected on machines from the Grid5000 infrastructure \cite{grid5000}, located in Rennes and Grenoble, France.
        \end{itemize}

        The data were then stored and processed on the University of Rennes' IT infrastructure.
        \\
        Data accessibility       & Repository name: Cheating in Networked Multiplayer Games: a Dataset. 
        
        Data identification number: \textit{to be updated — current status: draft on Zenodo}

        Direct URL to data: \cite{bertin_2026_19628775} \\
        Related research article &  None                                    \\ \hline
    \end{tabular}
    \caption{Specification Table}
    \label{tab:spec}
\end{table*}
\section{Value of the Data}\label{sec:valueData}

\begin{itemize}
    \item To the best of our knowledge, this is the first public dataset in which the use of a network flow manipulation cheat has been explicitly identified.
    \item The dataset contains traces of usage of diverse types of cheats: aimbot, wallhack, player-targeted DDoS, lag-switch, fixed-delay packet-tampering. 
    \item The explicit log of cheating actions enables the labelling of other gameplay event logs as either cheating or honest.
    \item The dataset includes traces at both the network and application levels, enabling analysis of network-level cheating across layers and the investigation of how events at one layer affect the other.
    \item The dataset contains data that was generated during real gaming sessions involving human players. Thus, it is representative of real-world gaming sessions involving cheaters.
    \item The data is intended for use by researchers studying cheat detection and gaming security, and can also be leveraged by industry and video games studio to develop and improve their cheats detection models. The dataset aims to contribute to mitigating cheating in online games.
\end{itemize}
\section{Background}\label{sec:Background}
The Multiplayer Online Games (MOGs) market was valued at US\$225 billion in 2025 \cite{fortune2025onlinegames}, yet it faces significant threats from cheating, which frustrates users, reduces engagement, and raises ethical concerns regarding fairness in e-sports \cite{johansson2025cheating, Granados2018bigproblem, apexLegendsEsportHacking}.

Cheats generally fall into two categories. The first manipulates the game client at runtime to access unauthorized information. For example, a \textit{wallhack} reveals hidden opponent locations (e.g., behind walls). This exposed information can be processed to automatically generate game inputs (e.g., an \textit{aimbot} automates aiming in shooter games \cite{choi2023botscreen}).

The second exploits network communications between game clients and the server:
\begin{itemize}
    \item \textit{Packet tampering}: cheats intercept and modify the content of the packets exchanged at the network level \cite{bertin2025disconnecting,laurens2007novel}. For example, a packet reporting a missed shot can be altered to communicate a successful hit.
    \item \textit{Disruption of packet flow}: Attackers manipulate network packet flow without changing packet contents. Examples include \textit{lag-switch}, \textit{fixed-delay} and \textit{player-targeted DDoS} attacks, which increase latency between clients and the server to disrupt game synchronization \cite{Benhabbour2023Attacks, Webb2007Cheating}. 
\end{itemize} 

Detecting network flow disruption cheats is challenging, as they could be confused with normal network latency from the server’s perspective. Whether such cheats can be detected remains an open question. Addressing this issue requires data; however, no dataset containing network traces of these cheats currently exists, motivating the creation of a dedicated dataset.

\section{Data Description}\label{sec:DataDescription}

The dataset contains logs of gaming sessions collected on a multiplayer online game. The dataset combines traces collected from both the game server and the clients (as described in Section \ref{sec:ExpDesign}). On each node, logs are collected at two levels: 1) at the game application layer, capturing the game events generated by the game logic and observable by the players, and 2) at the network level, collecting the data embedded in the network packets exchanged by the client and server. 

\subsection{Structure and Entities}

The dataset is organised under \textit{Sessions}, where each session represents a game session stored in its own folder identified by a unique session ID at \texttt{Sessions/Session\_id}. 
Each session follows a hierarchical structure composed of \textit{Scenarios}, \textit{Players}, and \textit{Match} containers, within which different types of data are stored. The content of these containers is described in the subsection hereafter.

% The dataset is organised into the following entities: \textit{Session}, \textit{Trace}, \textit{Capture}, \textit{Player}, and \textit{Scenario}. These entities are described in the subsections hereafter. Listing \ref{lst:dataset-tree} outputs the directory structure of the dataset. 

% Finally, \textit{Scenario} defines the experimental setup and is shared independently of sessions.

Listing \ref{lst:dataset-tree} illustrates the directory structure of the dataset.

\begin{lstlisting}[caption={Dataset directory structure}, label={lst:dataset-tree}]
dataset
`-- Sessions
    +-- Session_01
    |   +-- Scenarios
    |   |   +-- scenario_1.json
    |   |   +-- scenario_2.json
    |   |   `-- [...]
    |   +-- Players
    |   |   +-- player_A
    |   |   |   +-- sys_info.txt
    |   |   |   `-- feedback.md
    |   |   +-- player_B
    |   |   `-- [...]
    |   +-- Match_01
    |   |   +-- metadata.yml
    |   |   +-- Clients
    |   |   |   +-- trace_player_A_2025.12.08-11.57.35.bin
    |   |   |   +-- trace_player_B_2025.12.08-11.57.42.bin
    |   |   |   +-- traceroute_player_A.txt
    |   |   |   `-- traceroute_player_B.txt
    |   |   `-- Server
    |   |       +-- capture_2025.12.08-11.57.35.pcap
    |   |       `-- trace_2025.12.08-11.57.24.bin
    |   +-- Match_02
    |   |   `-- [...]
    |   `-- [Other Matches]
    +-- Session_02
    |   `-- [...]
    `-- [Other Sessions]
\end{lstlisting}

\subsubsection{Scenarios}
\textit{Scenario} is a formally defined gameplay description that can be interpreted by game clients. These scenarios were used in our experiment to automate game sessions and are included in the dataset to support the reproducibility of the experiments. The folder contains scenarios used in the related \textit{Session}.
\\

\subsubsection{Players}
\textit{Player} represents a participant taking part in the experiment. The data concerning a player is organized in a folder at the path: \texttt{Players/Player\_id}, where \texttt{Player\_id} is a unique identifier assigned to each participant to anonymize their name. The participant's environment information is stored in the text file named \texttt{sys\_info.txt}, which records the Operating System (Platform and Version), hardware information (CPU, number of cores, GPU, and RAM), and the Unreal Engine version. An example of \texttt{sys\_info.txt} file is shown in listing \ref{lst:env}. The folder also contains a form (\texttt{feedback.md}), filled out by the participants, presented in Appendix \ref{sec:feedback_form}. The form provides information about the players’ experience with online games and their geographical background, which could be leveraged to filter logs during analysis. 

\begin{lstlisting}[caption={Example of Environment file}, label={lst:env}]
    CPU=Intel(R) Core(TM) i5-10210U CPU @ 1.60GHz
    CPU Cores=4
    GPU=Intel(R) UHD Graphics
    Platform=Windows
    OS Version=10.0.19045.1.256.64bit
    Total RAM (GB)=8.0
    UE=d1e3064
    Plugin=03fea25
    Game=9eb8cbc
\end{lstlisting}

If players participate in multiple game sessions, their corresponding player folder is duplicated for each session. This allows session-specific feedback to be provided and enables changes in the system used to play.

\subsubsection{Match} The \textit{Match} directory represent one execution instance of the game in the \textit{Session}. The match recorded in the dataset can be found at \texttt{Sessions/Session\_id/Match\_id}, where \texttt{Match\_id} is a unique identifier for the session.

Each match contains a \texttt{metadata.yml} file that provides contextual information about the session recording. It specifies whether the session was conducted with real-world players or with automated game clients. The scenarios selected for automated game sessions are referenced in this file by a pointer to the \texttt{scenario\_id.json} listed under the \texttt{Sessions/Sessions\_id/Scenarios/} folder. Metadata also specifies the version of the environment used, identified by the commit hashes of the environment software (see Section \ref{sec:ExpDesign}). Additionally, it includes the list of cheats that were available for players in the game session. An example of a game session in Section \ref{sec:ExpDesign} lists a \texttt{metadata.yaml} file as an example.

The remaining content in the \textit{Match} folder consists of data collected from both clients and the server during gameplay. This data is organised into two folders: \textit{Clients} for client-side data and \textit{Server} for server-side data. This data can be separated into three different types of content: \textit{capture}, \textit{traceroute}, and \textit{trace}.

A \textit{Capture} file is a raw network packet capture of the packets exchanged between the clients and the server. The \textit{capture} is recorded only on the server-side, therefore stored in the \textit{Server} folder.

\textit{Traceroute} is a record of the evolution of the player's network environment through a history of \texttt{traceroute} command output (\texttt{tracert} on Windows) executed toward the server. Traceroute is executed only on the client side and is therefore stored in the \textit{Clients} folder. 

A \textit{trace} is a sequence of log entries that are recorded at runtime on both the clients and server. Each \textit{Match} contains one server trace and $N$ client traces, where $N$ is the total number of clients. Clients' traces are located in the \texttt{Clients} subfolder and the server trace is located in the \textit{Server} subfolder. 

The server \textit{traces} are timestamped using the match start time. Client traces are timestamped when each client joins the session. All the timestamps are expressed in UTC. Client traces refer to one \textit{player}, identified by its \texttt{Player\_id}. This information can be read in the file name of the \textit{trace}.

Log entries in a \textit{trace} can be of three types: \textit{GameEvent}, \textit{PacketLog}, or \textit{NetStat}. Each log entry contains two types of time information: a UTC timestamp (\textit{ts}) and a counter (\textit{frame}) that is updated every game loop iteration.
\\

\paragraph{Game events}

\begin{figure*}
    \centering
    \includegraphics[width=0.7\linewidth]{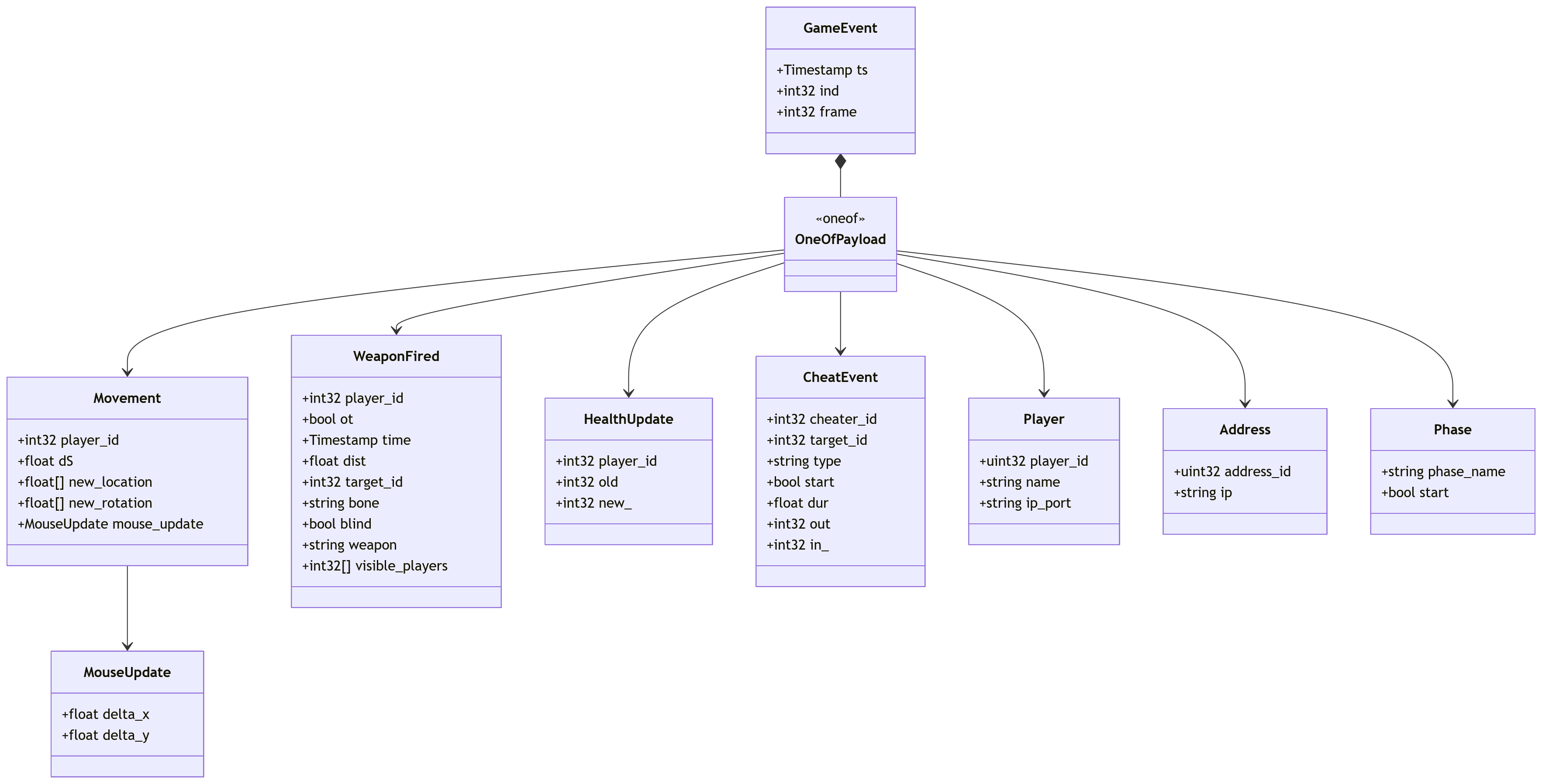}
    \caption{\textit{GameEvent} entry data structure.}
    \label{fig:gameEvents}
\end{figure*}

\textit{GameEvent} structures represent a log entry that has been generated at the game application level. This level of logs represents the events that are observable by the players. It is important to note that those events do not necessarily represent a direct translation of information received from the network but may also be events generated locally by the game logic. Figure \ref{fig:gameEvents} presents the data structure involved in the \textit{GameEvent}. Seven different types of events can be observed in the logs. \textit{Movement} captures the movements of the player's avatar. Client-side movements are combined with mouse movements (\textit{MouseUpdate}), which directly affect the player’s rotation. \textit{WeaponFired} records when a player fires a weapon, \textit{HealthUpdate} tracks changes in player's health, \textit{CheatEvent} logs the usage of cheats by a player. \textit{Player} identifies a player. \textit{Address} identifies the IP address used by a player to communicate with the server. \textit{Player} and \textit{Address} objects are created when a player joins the game session. \textit{Phase} indicates the timing of different gameplay match phases (e.g., warmup, gameplay). A comprehensive description of the data fields contained in \textit{GameEvents} is provided in Appendix \ref{app:GameEvents}.

The table \ref{tab:exampleGameEvent} presents an example of an event that has been extracted from the trace. We can read a \textit{weapon fired} event in which the player 4 fired a shotgun at player 2, hitting his right clavicle. The shot was on target (\textit{ot}) despite a distance of more than 2973 units of space. Players 2 and 4 refer to player identifiers that have been assigned, which are associated with player names recorded in the \textit{new\_player} events issued when they joined the game.
\\

\begin{table*}[htbp]
    \centering
    \scriptsize
    \caption{Example of a packet notification header extracted from a network-level log}
    \begin{tabular}{|c|c|c|c|c|c|}
    \hline
     type&timestamp & Sequence number & Acked Sequence & extra info & jit\_clock \\
    \hline
     notif\_hdr &2025-12-11 09:11:54.371 +00:00 & 8253 & 7698 & jit\_clk & 1023.0 \\
    \hline
    \end{tabular}
    \label{tab:exampleGameEvent}
\end{table*}

% \begin{table}[h!]
% \centering
% \caption{Specification of the \textit{GameEvent}: \textit{WeaponFired}}
% \label{tab:weaponfired}
% \begin{tabular}{lll}
% \toprule
% \textbf{Field} & \textbf{Type} & \textbf{Description} \\
% \midrule
% \textbf{WeaponFired}&&\\
% player\_id & int32 & ID of the player shooting. \\
% ot & bool & Whether shot targeted another player. \\
% time & Timestamp &  UTC-Timestamp (only if \textit{ot} true). \\
% dist & float & Distance to target (only if \textit{ot} true). \\
% target\_id & int32 & ID of player hit (only if \textit{ot} true). \\
% bone & string & Bone hit (only if \textit{ot} true). \\
% blind & bool & Whether shot was blind  (only if \textit{ot} true). \\
% weapon & string & Weapon used. \\
% \bottomrule
% \end{tabular}
% \end{table}

\begin{figure*}
    \centering
    \includegraphics[width=\linewidth]{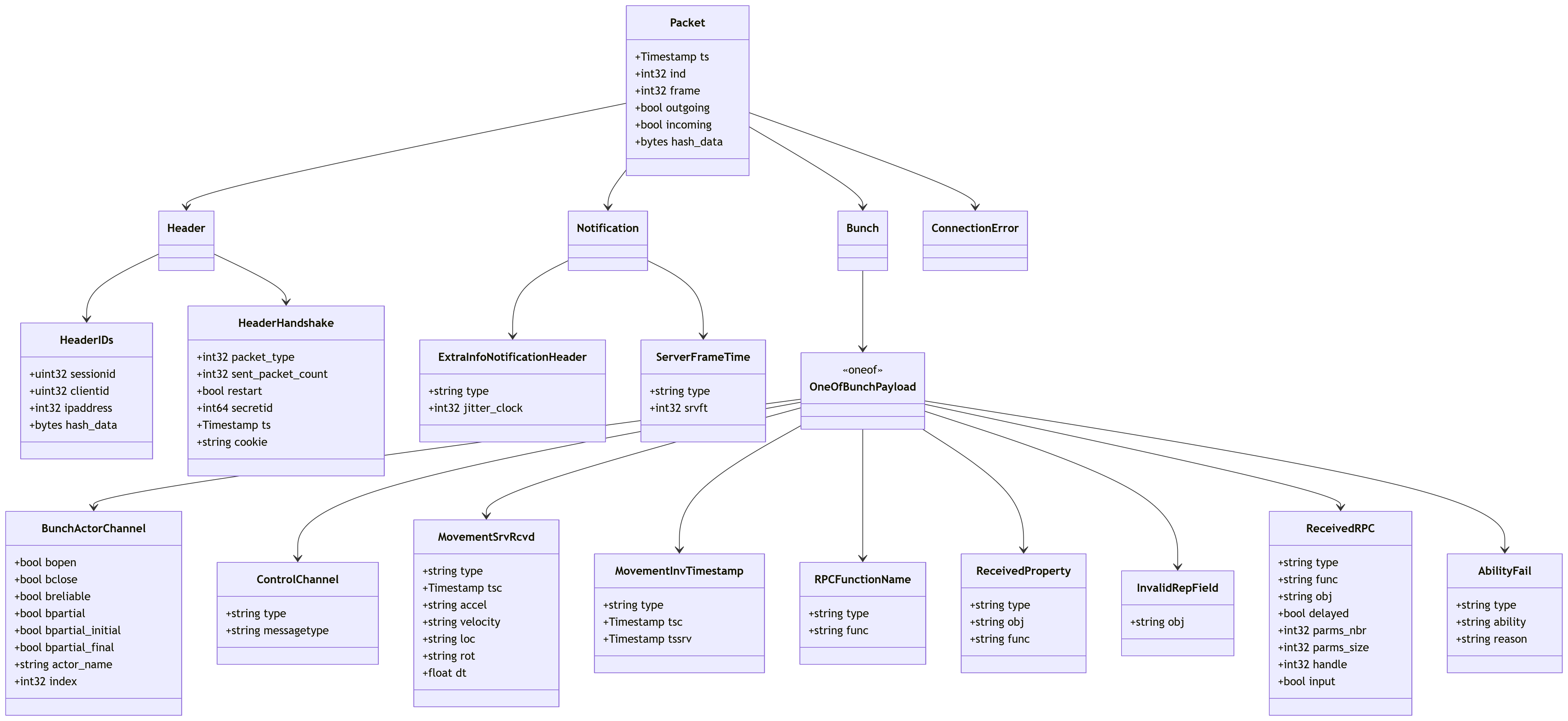}
    \caption{\textit{PacketLog} entry data structure.}
    \label{fig:packets}
\end{figure*}

\paragraph{Packets logs}
\textit{Packet logs} are logs collected by the game engine upon network packets transmission and reception. They record different fields from the Unreal Engine networking protocol \cite{bertin2025disconnecting}. Figure \ref{fig:packets} presents the structure of those logs. The log of a \textit{packet} is organised under four categories of information, which represent the different fields in the UE application-layer protocol: \textit{Header}, \textit{NotificationHeader}, \textit{Bunches}. An additional category has been added to represent the potential connection error messages that may be sent. A comprehensive description of the data fields contained in \textit{PacketLogs} is provided in Appendix \ref{app:packetLogs}.

 Table \ref{tab:examplePacketLogs} presents an example of a notification header extracted from a packet logged in the trace. The header includes the timestamp corresponding to the packet's reception, as well as the Sequence and Acknowledged Sequence numbers of the packet from Unreal Engine's networking protocols. This notification header also embeds additional information, including a jitter clock value, used by the engine for network condition measurement.
 \\

\begin{table*}[htbp]
    \centering
    \scriptsize
    \caption{Example of a weapon-fired event}
    \begin{tabular}{|c|c|c|c|c|c|c|c|}
    \hline
     event\_type&timestamp & player\_id & weapon & ot & dist & target\_id & bone \\
    \hline
     weapon&2025-12-11 09:16:38.658 +00:00 & 4 & B\_Shotgun\_C\_2147476661 & True & 2973.068848 & 2 & clavicle\_r \\
    \hline
    \end{tabular}
    \label{tab:examplePacketLogs}
\end{table*}

\paragraph{Network Statistics}
\textit{NetStats} is a log entry that provides information about client-server latency. It is only present in the \textit{client traces} and periodically records the evolution of the latency to reach the server (1s). This data structure contains two values: \textit{ICMPPing} and \textit{UDPPing}, which represent latency measurements respectively obtained via ICMP and UDP, as described in Section \ref{sec:ExpDesign}. Additionally, the latency of each packet transmission can be retrieved by combining sender and receiver logs and comparing their timestamps. The machines used in this dataset were synchronized using NTP.

\subsection{Data Labelling}

The dataset contains logs recorded on an online game, including the activation and deactivation of cheats. This information can be used to label logs that correspond to either honest or cheating behaviour. Since all recorded logs are timestamped, cheating labels can be assigned to the logs concerning the cheater's actions, timestamped in the cheat activation time interval. In the case of a player-targeted DDoS cheat, the effects of the cheat primarily target the victim’s gameplay rather than the cheating client (as detailed in \ref{subsubsec:onlinegame}). Thus, a victim label could be applied to the target's logs. For instance, based on the timestamp of the example of log describing a lag-switch event in Table \ref{tab:exampleCheat}, we can infer that the shoot event presented in Table \ref{tab:gameevents} was performed by player 4 while cheating. 

\begin{table*}[htbp]
    \centering
    \caption{Example of a cheat event}
    \scriptsize
    \begin{tabular}{|c|c|c|c|c|c|c|c|}
    \hline
     event\_type&timestamp & cheater\_id & type & start & duration \\
    \hline
         cheat&2025-12-11 09:16:38.460 
         
         +00:00 & 4 & Lag-Switch & True & 0.5 \\
    \hline
    \end{tabular}
    \label{tab:exampleCheat}
\end{table*}

\subsection{Data Format}\label{sec:format}
The entire dataset is compressed using gzip. \textit{Captures} are packet capture files (\texttt{.pcap}), containing raw packets. \textit{Traces} are stored in a binary format, consisting of serialised data following a specification defined in a Protocol Buffer schema \cite{google2025protobuf}. The process of serialization has been realized through the Protocol Buffer engine and can be leveraged to deserialise the data in order to analyse it.
We distribute the specification file (\texttt{.proto}) to this purpose. As protobuf is language-agnostic, this facilitates the manipulation of the data in various programming languages. In order to facilitate the process of extraction, we also distribute a parser developed in Python, that translates a trace as input to pandas dataframes or to a CSV file, available on the following Git repository \footnote{\url{https://github.com/Lyra-FullGame/ParserProtobufLogs}}. 

We describe in the Appendix \ref{tab:top-levelstructs}, the different data structures defined in this \texttt{.proto} specifications used to organize the logs and that can be leveraged for data analysis if not using our Python parser tool, but by directly rewriting another parser.

% \subsubsection{Top-level structures}
% This paragraph describes the data structures defined in the \texttt{.proto} specification. 
% Top-level structures are designed to organize log entries during serialization and deserialization. They are abstracted when using the parser provided, but we describe them here for anyone who wishes to implement a custom parser. \textit{RingBufferLog} is the buffer used for storage; any entry of this buffer is a \textit{LogEntry}. \textit{LogEntry} contains one of the two following data structures: \textit{GameEvent} or \textit{Packet}. 

% Appendix \ref{app:TopLevel} presents all the entries that can be found in the various data structures used to represent our logs. \textit{Field} refers to the name of the field in the schema, \textit{Type} to the type used to represent the data, and \textit{Description} describes the meaning of the data.  

% \subsection{Statistics}
% \hb{Add statistics: Total size in GB (compressed or not), Total number of traces, total number of players. Size of a client/server trace: mean, max, min. Number of Log Entry types/ trace: PacketLog, GameEvent.}\hb{To update after finishing to compile the dataset}

\section{Experimental Design, Materials and Methods}\label{sec:ExpDesign}

\subsection{Experimental environment}\label{subsec:ExpEnv}

The dataset was collected using a multiplayer online game integrated with a framework that implements the cheats and enables data collection. 
\\

\subsubsection{Online Game}\label{subsubsec:onlinegame}
The game is a multiplayer networked game developed with the Unreal Engine (UE), one of the leading game engines of the market \cite{2025gameengines}. Thus, the game used in the experiments shares pieces of software with other UE titles such as Fortnite, PUBG, and Valorant.

The game is a multiplayer shooter game based on the  \textit{Lyra Starter Game} UE template \cite{lyra2025steam}. Players compete in a team deathmatch mode, where two teams score points by eliminating players from the opposing team. The network architecture is client-server. The changes made to the template aim to incorporate the cheats directly into the game design. The cheats can be viewed as specific abilities that are periodically available to the player. The player picks one of the following cheats as a main ability: \textit{Lag-Switch}, \textit{player-Targeted DDoS}, and \textit{Fixed-Delay}. The other cheats can be picked up as bonuses on the game map: \textit{Aimbot}, \textit{HitRedirection} (Packet-Tampering cheat), and \textit{Wallhack}. The abilities and pick-ups are available to every player of the game. Reloading times are applied between consecutive usages of an ability and extra pick-ups. This strategy ensures that the logs contain both honest and cheater records for all the players. These mechanics have also been selected for ethical concerns, avoiding the promotion of cheating when it is not allowed by the game rules.

It is important to note that for ethical reasons, the \textit{Player-targeted DDoS} attack is not actually performed but instead simulated within the framework, as it is a real network attack that could have adverse effects on the player's network. To simulate the attack, the game server tells the game client to drop packets according to configurable packet loss rates for both incoming and outgoing traffic flows, which are configurable. This configuration is logged (see \textit{CheatEvent} in Appendix~\ref{app:GameEvents})
\\

\subsubsection{Log Collection framework}
\begin{figure*}[t]    
    \centering
    \includegraphics[width=0.8\linewidth]{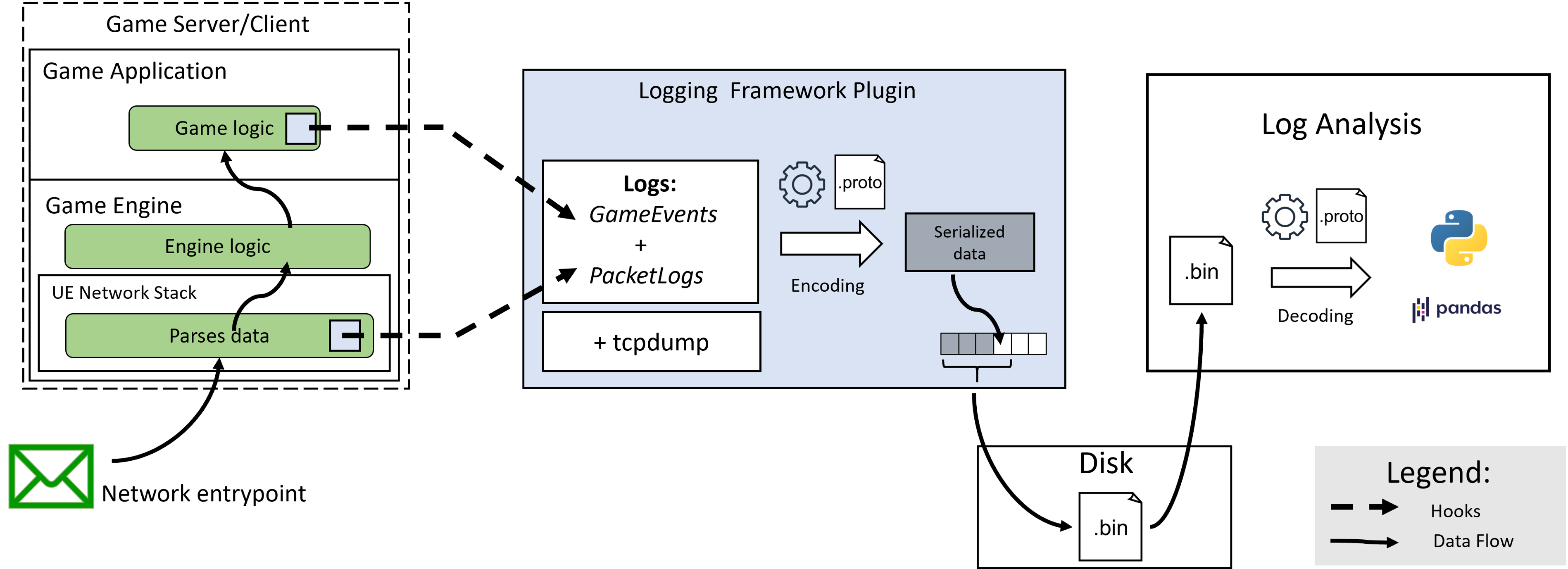}
    \caption{Log collection pipeline}
    \label{fig:logArch}
\end{figure*}

Log collection and cheat detection are implemented within a game engine plug-in, based on a framework designed to enable data collection on cheating in online game environments \cite{shaikh2025eicc}. The framework contains different modules. The first module implements and enables cheating directly from within the game. By doing so, it provides a view into players' use of cheats and provides a way to reproduce cheating without relying on external tools. The cheats were implemented to replicate the same mechanics and effects as those observed when real cheats are used in actual Unreal Engine games. The second module collects logs from the gaming session following the specification presented in Section \ref{sec:DataDescription}. As represented in Figure \ref{fig:logArch}, these logs are collected at two different levels in the game engine. 

The first level corresponds to the game application layer, where events are processed, and game logic is executed. This represents the state of the game and includes the events perceived by the players. 

The second level is the engine network stack, where the applicative data received from network communication is parsed. This level focuses on filtering and decoding the embedded application payload to extract fields defined by Unreal Engine network protocols, while also capturing timing information. Thereby providing insights into both the information exchanged and the temporal behaviour of online game communication. This represents a vision of online game communication at the network level. 

To complete the capture of network communications, we additionally collect raw network traffic at the server level. For this purpose, we use the \texttt{tcpdump} tool, which captures transmitted and received packets and stores them in a \texttt{pcap} file. We apply the following filter to the \texttt{tcpdump} capture: \texttt{udp and port \${port}}, where \texttt{port} denotes the game server’s port.
% This is only performed at the server level for technical reasons, as it would require third-party installation on clients and elevated Windows administrative privileges. \hb{Discuss this}

To complete the collection of network information, we periodically collect from the client, the latency to reach the server (each second). This is done by combining the results of ICMP ping to the server with UDP-ping already implemented in Unreal Engine. In addition, the \texttt{traceroute} command (or \texttt{tracert} on Windows) is performed at the beginning of each client session to obtain detailed information about the player's network infrastructure. When a spike of latency is detected, the command is executed again to observe potential network path changes. 
% The latency is collected every second. When a spike of latency is detected, a new \texttt{traceroute} is performed to detect an infrastructure evolution. 

\subsection{Scripted Game Session Examples}
A subset of the game session included in the dataset was recorded as "scripted-examples", which reproduce specific, expected game scenarios that contain a low level of noise, much lower than the one produced by less deterministic human gameplay. Scripted-Example traces can be divided into two different sets: \textit{fully-automated} and \textit{human-executed reference} traces.
\\

\subsubsection{Fully-Automated Traces}\label{sec:fully-automated}
% A subset of the game sessions in the dataset was collected while automating game scenarios at the clients. This is useful to reproduce specific gameplay scenarios while varying the environment parameters. 
% An example available in the dataset reproduces the same gameplay scenario while changing the network topology and the latency between the server and clients. 
Fully-Automated game sessions were collected while automating game scenarios at the clients.
For this purpose, we leveraged the automation module included in the log collection framework \cite{shaikh2025eicc}. The module automates gameplay activities, such as movements, weapon shots, and cheat usage. Specific gameplay scenarios can be scripted using a \texttt{JSON} file that defines the actions to automate and their parameters. 
% The language specification used to describe the automated actions, such as movements, shots, and cheat usage, is available on the plugin repository \hb{See if we can do something here, otherwise remove}.

We insert 3 game sessions in the dataset that were recorded with automated scenarios, which serve as examples for analysis with known ground-truth scenarios. The experimental setup combines a game server located at Gravelines, France, and two game clients at two different locations in France: Grenoble and Rennes. All the machines run on Linux. This information can be found in the dataset besides the hardware information in the \texttt{env.txt} file for the clients and in the \texttt{metadata.yml} for the clients. 
% The average network latency between the clients and the server is shown in Table \hb{insert}. This information is observable in the dataset at \texttt{Latency/latency.txt}. 
We describe hereafter the different scenarios that were followed. Each scenario was recorded on a session of 10 minutes, and reproduced 10 times.
\\

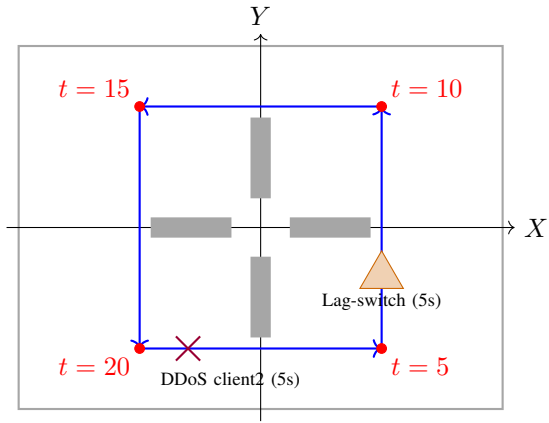
\begin{figure}[h]
\centering
\begin{tikzpicture}[x=0.8cm,y=0.8cm]

    % ------------------
    % Colors
    % ------------------
    \colorlet{mapcolor}{gray!70}
    \colorlet{pathcolor}{blue}
    \colorlet{pointcolor}{red}

    % ------------------
    % Map boundary
    % ------------------
    \draw[thick, mapcolor] (-4,-3) rectangle (4,3);

    % ------------------
    % Axes
    % ------------------
    \draw[->, black] (-4.2,0) -- (4.2,0) node[right] {$X$};
    \draw[->, black] (0,-3.2) -- (0,3.2) node[above] {$Y$};

    % ------------------
    % Walls
    % ------------------
    \draw[thick, mapcolor, fill] (-0.15,0.5) rectangle (0.15,1.8);
    \draw[thick, mapcolor, fill] (-0.15,-1.8) rectangle (0.15,-0.5);
    \draw[thick, mapcolor, fill] (-1.8,-0.15) rectangle (-0.5,0.15);
    \draw[thick, mapcolor, fill] (0.5,-0.15) rectangle (1.8,0.15);

    % ------------------
    % Player trajectory
    % ------------------
    \draw[thick, ->, pathcolor] (2,-2) -- (2,2);
    \draw[thick, ->, pathcolor] (2,2) -- (-2,2);
    \draw[thick, ->, pathcolor] (-2,2) -- (-2,-2);
    \draw[thick, ->, pathcolor] (-2,-2) -- (2,-2);

    % ------------------
    % Stop points
    % ------------------
    \filldraw[pointcolor] (2,-2) circle (0.08)
        node[below right] {$t=5$};
    \filldraw[pointcolor] (2,2) circle (0.08)
        node[above right] {$t=10$};
    \filldraw[pointcolor] (-2,2) circle (0.08)
        node[above left] {$t=15$};
    \filldraw[pointcolor] (-2,-2) circle (0.08)
        node[below left] {$t=20$};

    % % ------------------
    % % Player
    % % ------------------
    % \begin{scope}[yshift=1.2cm,xshift=0.1cm]
    %     \draw[black] (2.6,-2.2) circle (0.12);
    %     \draw[black] (2.6,-2.32) -- (2.6,-2.7);
    %     \draw[black] (2.5,-2.45) -- (2.7,-2.45);
    %     \draw[black] (2.6,-2.7) -- (2.45,-2.9);
    %     \draw[black] (2.6,-2.7) -- (2.75,-2.9);
    
    %     \node at (2.6,-3.05) {Player};
    % \end{scope}

    \node[cheatlag] at (2,-0.8) {};
    \node[below=3pt] at (2,-0.8) {\scriptsize Lag-switch (5s)};
    
    \draw[cheatdos] (-0.7,-1.1) -- (-0.5,-0.9);
    \draw[cheatdos] (-0.5,-1.1) -- (-0.7,-0.9);
    \node[below=3pt] at (-0.5,-2.1) {\scriptsize DDoS client2 (5s)};

\end{tikzpicture}
\caption{Automated Scenario for client 1. Game map with walls (grey), coordinate axes (black), and scripted player movement (blue) with stop points (red). Lag-Switch ($\triangle$) and DoS (\texttimes) targeted against client 2 are triggered for 5 seconds, at $t=7$ and $t=23$.}
\label{fig:scenario1}
\end{figure}

\paragraph{Scenario 1}
The purpose of this first series of automated sessions is to observe the effects of Lag Switch and player-targeted DDoS on the movement actions in the game. For this purpose, we scripted a scenario for each client. The scenario for client 1 is displayed in Figure \ref{fig:scenario1}. It is a scenario involving simple movements that follow a square pattern, with usage of the Lag-Switch cheat at the timestamp $t=5s$, and triggers a DOS against the second player at $t=25s$. The definition of a movement is done through the plugin scripting language and repeated multiple times. Example of movement and cheat scripting is provided in the Listing \ref{lst:scripting}. The second client is automated with movements similar to client 1, with a delay of 5 seconds. This scenario is repeated every 35 seconds. 
\\

\begin{lstlisting}[caption={Game scenario scripting with three types of actions.
\textit{Move} defines a movement to coordinates $(X,Y)$, bounded by a specified duration (5\,s in this example).
\textit{Cheat} defines a cheating action, here a denial-of-service (DoS) attack targeting the player \texttt{PTB*} for 5\,s.
\textit{Shoot} defines a shooting action targeting player \texttt{PTB*}, or the coordinates $(X,Y,Z)$ if the target is not visible.
The symbol \texttt{*} denotes a wildcard in the player's full name.}, label={lst:scripting}]
    {   "Action": "Move",
        "Timestamp":5.0,
        "Duration":5,
        "X":2000,        
        "Y":-2000   
    },{
        "Action": "Cheat",
        "CheatType": "DOS",
        "Timestamp": 24.0,
        "Duration": 5.0,
        "Target": "PTB*" 
    },{   
       "Action": "Shoot",
        "Timestamp": 25.0,
        "X": 670,
        "Y": 720,
        "Z": 100,
        "Target": "PTB*"    }
\end{lstlisting}

\paragraph{Scenario 2}\label{par:scenario2}
The second automated scenario builds on the first scenario by adding weapon fire events combined with cheating actions. Shoots are performed by client 1 on client 2. They are performed 0.5 seconds after client 1 cheats: first Lag-Switch, then DoS (similar to that presented in Figure \ref{fig:scenario1}). Shoots can be automated thanks to the \textit{Shoot} events in Listing \ref{lst:scripting}. The player aims at the targeted player when it is visible; otherwise, the shooter aims to the specified coordinates.
\\ 

\begin{figure}[h]
\centering
\begin{tikzpicture}[scale=0.7]

    % Colors
    \colorlet{mapcolor}{gray!70}
    \colorlet{pathcolor}{blue}
    \colorlet{pointcolor}{red}
    \colorlet{pointcolor2}{green!60!black}

    % Axes
    \draw[->] (-2.5,0) -- (2.5,0) node[right] {$X$};
    \draw[->] (0,-2.2) -- (0,2.2) node[above] {$Y$};

    % Walls
    \draw[thick, mapcolor, fill] (-0.15,0.5) rectangle (0.15,1.8);
    \draw[thick, mapcolor, fill] (-0.15,-1.8) rectangle (0.15,-0.5);
    \draw[thick, mapcolor, fill] (-1.8,-0.15) rectangle (-0.5,0.15);
    \draw[thick, mapcolor, fill] (0.5,-0.15) rectangle (1.8,0.15);

    % Points
    \filldraw[pointcolor] (-1.3,1.3) circle (0.15)
        node[above] {client 2};
    \filldraw[pointcolor2] (-1.3,-1.3) circle (0.15)
        node[below] {client 1};

\end{tikzpicture}
\caption{Initial settings ($t=0s$) for clients 1 and 2 in scenario 3, with map walls (gray). Coordinates axes intersect in the center of the map.}
\label{fig:scenario3}
\end{figure}
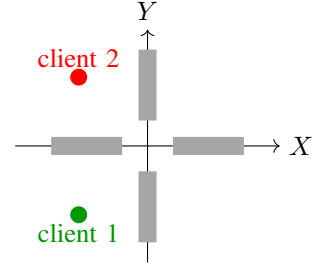

\paragraph{Scenario 3}
The third scenario involves two clients who are cheating. The movements performed by the clients are described in Figure \ref{fig:scenario3}. At the beginning, both of the clients are behind walls, and cannot see each other. Then, in turn, they move to the centre of the map, where they are uncovered. First, client 1 moves and shoots at client 2, who is stationary. Then client 1 comes back to its previous location to be covered by the walls. At $t=10s$, client 2 moves to the centre and, client 1 shoots at client 2 as soon as it becomes visible. This scenario is then restarted with client 1 cheating. Client 1 first uses a lag switch while moving towards the centre. Secondly, client 1 launch a DDoS attack against client 2, while client 2 is moving. Then Client 1 uses a Fixed-Delay while moving to the centre.

For each automated session in the dataset, the scenario script file is provided. 
\\

\subsubsection{Human-Executed Reference Traces}\label{sec:humanRef}
Human-executed reference traces correspond to scripted gameplay scenarios recorded by a human player. The dataset contains three sets of three traces, recorded with two clients on the same set-up used in fully-automated traces (§\ref{sec:fully-automated}). 

Within each set of traces, the first trace capture the human player defeating its opponent without using any cheat; the second trace the same outcome with the help of the lag-switch, the last one records the outcome with the help of the DDoS cheat. The series differ in the opponent actions. In the first one, the opponent is neither moving nor shooting; in the second the opponent is moving, following the Scenario 2 presented in §\ref{sec:fully-automated}. In the third one the opponent shoots at the player when he can see it.

\subsection{Experimental campaign}\label{section_realhuman}
This subsection presents the experimental protocol conducted to collect traces involving real players who participated in gameplay sessions.
\\

\subsubsection{Participants}
The experimental campaigns are conducted with real players, who were recruited through academic networks. All the participants gave their consent to participate in the experience. The screened participants represented a wide range of experience levels (labelled by player in the dataset). Participants are required to have access to a computer running one of the following OSes: Linux, Windows, MacOS; and the minimal hardware requirements that can be found for Lyra on Steam \cite{lyra2025steam}. A summary of the sessions realized is presented in Table \ref{tab:SessionsSummary}, showing the number of players, the diversity of geographical locations, which may influence the network conditions, and their experience, which may influence their gameplay behaviours.
\\

\begin{table*}[h!]
    \centering
    \begin{tabular}{|l|c|c|c|l|l|}
    \hline
    \textbf{id} & \textbf{\# Players} & \textbf{\# Automated} & \textbf{\# Matches} & \textbf{Clients Locations} & \textbf{Description} \\
    \hline
    01 & 6 & 0 & 4 & Benin, Saudi, France & Gameplay with real participants (see \ref{section_realhuman}) \\
    02 & 6 & 0 & 5 & Brasil, France, HongKong & Gameplay with real participants (see \ref{section_realhuman}) \\   
    03 & 6 & 0 & 11 & Cameroun & Gameplay with real participants (see \ref{section_realhuman})\\
    04 & 4 & 0 & 6 & Cameroun & Gameplay with real participants (see \ref{section_realhuman})\\
    05 & 0 & 2 & 5 & France (In-Lab) & Fully-automated traces, Scenario 1 (see \ref{sec:fully-automated})\\
    06 & 0 & 2 & 5 & France (In-Lab) & Fully-automated traces, Scenario 2 (see \ref{sec:fully-automated})\\
    07 & 0 & 2 & 5 & France (In-Lab) & Fully-automated traces, Scenario 3 (see \ref{sec:fully-automated})\\
    08 & 1 & 1 & 4 & France (In-Lab) & Human-Executed Reference traces (See \ref{sec:humanRef})\\
    09 & 1 & 1 & 4 & France (In-Lab) & Human-Executed Reference traces (See \ref{sec:humanRef})\\
    10 & 1 & 1 & 4 & France (In-Lab) & Human-Executed Reference traces (See \ref{sec:humanRef})\\
    \hline
    \end{tabular}
    \caption{Summary of the sessions and player groups in the dataset. \emph{Session\_id} denotes the session identifier, \textit{\# Players} indicates the number of human players, \textit{\# Players} indicates the number of automated clients, \textit{\# Matches} the number of matches played, \textit{Locations} lists the countries where the players were located, and \textit{Descripion} provides additional details about the game session context.}
    \label{tab:SessionsSummary}
\end{table*}

\subsubsection{Experimental protocol for a game session}

When a group of participants is identified, we carry out the following procedure.

\paragraph{Before the game session} We provide the participants with the game client software and setup instructions for installation on their computer. A specific version of the binary is available for each OS. The binary to install is an executable package containing the whole client's experimental environment (game client and logging framework).
We also provide video tutorials explaining the game mechanics and the consequences of the cheats on the gameplay.

\paragraph{The game session} After verifying that all the players can launch the game and automatically connect to the game server, a game session was started with all participants playing a team deathmatch. During the game, the logs described in the dataset specification (Section \ref{sec:DataDescription}) are collected. The duration of a match is 10 minutes. The number of players by session is 6 to 8, divided into two teams.

\paragraph{After the game session} Participants fill the \texttt{feedback.md} form to detail their experience relative to the game. The detailed questions are provided in Appendix \ref{sec:feedback_form}. Then we asked them to upload to a repository a compressed folder containing the traces generated, the game logs, and the feedback document filled out. 
The framework and the binaries are available to researchers who are interested in our works.
\\

\subsubsection{Example of Session in the Dataset}
In this subsection, we detail the conditions under which the \texttt{Session\_01} was conducted, taking the \texttt{metadata.yml} file included in the dataset as an example. The session involved six players. The shortened commit hashes of the software used in the experimental environment are provided. Automation was disabled, as the session was realized with real players. The session lasted 27.5 minutes. Additionally, the hardware and system specifications of the game server, along with the cheats available in the game, are documented.

\begin{lstlisting}[caption={Metadata file describing Session\_01}, label={lst:metadata.yml}]
    clients_number: 6

    env:
      UE: d1e3064
      Plugin: 03fea25
      Game: 9eb8cbc
    
    automation: no
    start_time: 2026.01.14-19.52.56:337
    end_time: 2026.01.14-20.20.27:881
    
    server:
      city: Gravelines
      country: France
      hardware:
        cpu:Intel Core Processor (Haswell, no TSX)
        cpu_cores: 6
        ram: 12.0
      os:
        platform: Linux
        os_version: ubuntu25.04
    
    cheats: [DOS, LS, Fixed-Delay, Aimbot, Wallhack, HitRedirection]
\end{lstlisting}

The information relative to the player environment and experience can be fetched under \texttt{Players/Player\_id}. For instance, the \texttt{Player\_A}, included in this session was located in Rennes, France, and was experienced with shooter games. This can be observed by its CS:GO ranking provided in the \texttt{feedback.md}. Additionally, as shown in the \texttt{env.txt}, the player was on Linux.
\section{Limitations \textit{(limit: 200 words)}}\label{sec:Limitations}

The dataset focuses solely on a limited number of cheats that we have implemented in an execution sandbox to reproduce similar effects to real cheats. However, the experimental framework, and consequently the dataset, can be enriched with new cheats. Any newly added cheats will be recorded in the metadata file for the corresponding sessions.

The cheats used to collect these logs were implemented manually within the game, and are thus not the exact software used by actual cheaters. Nevertheless, they implement the same purpose conceptually and exhibit behaviours similar to those of real cheat software.

The cheats provided to the player are accessible through the game interface according to the game rules defined. Thus, users do not have full autonomy in tuning the cheat usage parameters. Nevertheless, the remaining flexibility in the cheat usage (timing, execution) corresponds to real-world cheating behaviour.

Additionally, the logs were collected from a single game, which may differs with other ones in terms of gameplay. However, since the dataset is based on the Unreal Engine, it benefits from sharing many common components with other games built on the same engine.
\section{Ethic Statement}\label{sec:ethic}
All the participants in the experiment gave their consent to the collection of their IP addresses for the purpose of compiling this dataset through the feedback form they filled.

\section{Credit Author Statements}\label{sec:credit}
% \hb{Guidelines: outline the contributions of each co-author, using the categories listed on this webpage \url{https://www.elsevier.com/researcher/author/policies-and-guidelines/credit-author-statement}}

\textbf{Hugo Bertin:} Conceptualization, Methodology, Software, Data Curation, Investigation, Writing - Original Draft\\
\textbf{Marc Dacier:} Conceptualization, Methodology, Writing - Review \& Editing, Supervision, Funding Acquisition.\\
\textbf{Yérom-David Bromberg:} Conceptualization, Methodology, Writing - Review \& Editing, Supervision, Funding Acquisition.

\section{Acknowledgements}\label{sec:ack}
The authors thank Salman Shaikh for his participation in the development of the software used in these experiments and in the organisation of the experimental gameplay sessions.  
  
The authors would also like to thank Cnam-Enjmin, its director Axel Buendia, and the students who contributed to the development of the online game, Timothée Drugeon, Samuel Suzan, Baptiste Denis, and Rémi Bernard, for their valuable work on the experimental platform used to collect this dataset.

\section{Declaration of competing interests}\label{sec:interest}
The authors declare that they have no known competing financial interests or personal relationships that could have appeared to influence the work reported in this paper.

%% Add \usepackage{lineno} before \begin{document} and uncomment 
%% following line to enable line numbers
%% \linenumbers

%% main text
%%

%% The Appendices part is started with the command \appendix;
%% appendix sections are then done as normal sections
% \appendix
% \section{Example Appendix Section}
% \label{app1}

% Appendix text.

%% For citations use: 
%%       \citet{<label>} ==> Lamport (1994)
%%       \citep{<label>} ==> (Lamport, 1994)
%%
% Example citation, See \citet{lamport94}.

%% If you have bib database file and want bibtex to generate the
%% bibitems, please use
%%
%%  \bibliographystyle{elsarticle-harv} 
%%  \bibliography{<your bibdatabase>}

%% else use the following coding to input the bibitems directly in the
%% TeX file.

% \bibliographystyle{ieeetr}  
% \bibliography{references}

\bibliographystyle{IEEEtran}
\bibliography{references}

% \printbibliography
\appendix
\section{Data Specification}
\onecolumn
\subsection{Top-Level Structures and global identifiers}\label{app:TopLevel}

Top-level structures are defined in the \texttt{.proto} specification, to organize log entries during serialization and deserialization. They are abstracted when using the parser provided, but we describe them here for anyone who wishes to implement a custom parser. \textit{RingBufferLog} is the buffer used for storage; any entry of this buffer is a \textit{LogEntry}. \textit{Player} is a data structure leveraged to identify players. Both \textit{PacketLog} and \textit{GameEvent} refer to \textit{Player} in their content. \textit{LogEntry} contains one of the two following data structures: \textit{GameEvent} or \textit{PacketLog}. Messages are structured data format defined in the Protobuf schema.

\begin{longtable}{p{3.5cm} p{3cm} p{8cm}}

\toprule
\textbf{Field} & \textbf{Type} & \textbf{Description} \\
\midrule
\endhead

% ======================= RingBufferLog =======================
\multicolumn{3}{l}{\textbf{Message: RingBufferLog}} \\
capacity & int32 & Maximum number of log entries stored. \\
head & int32 & Index of next entry to read. \\
tail & int32 & Index of next entry to write. \\
entries & LogEntry[] & Circular buffer content. \\[4pt]

% ======================= LogEntry =======================
\multicolumn{3}{l}{\textbf{Message: LogEntry}} \\
entry & oneof\{ \\
packet & PacketLog & Log entry containing a \textit{PacketLog} \\
event & GameEvent & Log entry containing a \textit{GameEvent} \\
event & NetworkStat & Log entry containing a \textit{NetStat} \\
\} \\[4pt]

% ======================= Player =======================
\multicolumn{3}{l}{\textbf{Message: Player}} \\
id & uint32 & Unique player identifier. \\
name & string & Player name. \\[4pt]

\bottomrule
\label{tab:top-levelstructs}
\end{longtable}

\subsection{Game Events}\label{app:GameEvents}

\begin{longtable}{p{3.5cm} p{3cm} p{8cm}}

\toprule
\textbf{Field} & \textbf{Type} & \textbf{Description} \\
\midrule
\endhead

% ======================= GameEvent =======================
\multicolumn{3}{l}{\textbf{Message: GameEvent}} \\
payload & \textit{oneof} \{ \\
movement & Movement & Player movement event. \\
weapon & WeaponFired & Weapon firing event. \\
health & HealthUpdate & Health update event. \\
cheat & CheatEvent & Cheat usage event. \\
new\_player & Player & Player join event (useful for identifying players at game-level analysis). \\
new\_address & Address & New IP address connection (useful for packet-level analysis). \\
phase\_event & Phase &  Phase of the Match (Warmup, Playing or PostGame) \\
\} \\
ts & Timestamp & Event timestamp. \\
ind & int32 & Index in ring buffer. \\
frame & int32 & Frame counter (internal engine counter, incremented every game loop). \\[4pt]

% ======================= Phase =======================
\multicolumn{3}{l}{\textbf{Message: Phase}} \\
phase\_name & string & Warmup, Playing, PostGame. \\
start & bool & True: beginning of the phase, False: end. \\[4pt]

% ======================= Address =======================
\multicolumn{3}{l}{\textbf{Message: Address}} \\
address\_id & uint32 & unique ID of the ip address(different from the player\_id). \\
ip & string & IP address and port number of the connection ("IP:Port") \\[4pt]

% ======================= Player =======================
\multicolumn{3}{l}{\textbf{Message: Player}} \\
player\_id & uint32 & unique ID of the player (different from the address\_id). \\
string & name & Name of the player (as recorded by UE). \\
string & address & IP address and port used by the player when joining the game ("IP:Port"). Warning: the source port of a client connection can change due to NATing during the game. In this scenario, UE updates the connection without modifying the player's object. Therefore, the port stored can be outdated. In that scenario the Session\_id and Client\_id in the \texttt{Packet::HeaderIDs} do not change. \\[4pt]

% ======================= Movement =======================
\multicolumn{3}{l}{\textbf{Message: Movement}} \\
player\_id & int32 & ID of the moving player. \\
dS & float & Time delta with previous Movement. \\
location & float[3] & Location (x,y,z) of the player's avatar. \\
rotation & float[3] (opt) & rotation (Pitch, Yaw, Roll) of the player's camera (only logged on server and owning client). \\
ts\_client & float & Client timestamp (seconds since the start of the game); only logged on owning client (movement of the avatar owned by the client) \\
mouse\_update & MouseUpdate (opt) & Mouse movement delta (Only logged client-side)  \\[4pt]

% ======================= Mouse Update ====================
\multicolumn{3}{l}{\textbf{Message: MouseUpdate}} \\
delta\_x & float & Mouse delta on the \textit{x}-axis (Pitch). \\
delta\_y & float & Mouse delta on the \textit{y}-axis (-Yaw). \\[4pt]

% ======================= WeaponFired =======================
\multicolumn{3}{l}{\textbf{Message: WeaponFired}} \\
player\_id & int32 & ID of the player shooting. \\
weapon & string & Weapon name. \\
ot & bool & Whether the shot reached an opponent (on target). \\
time & float &  Duration of the shoot (only if \textit{ot} true). \\
dist & float & Distance to target (only if \textit{ot} true). \\
cartridge\_id & int32 & Unique ID for the shoot (computed randomly). \\
target\_id & int32 & ID of player hit (only if \textit{ot} true). \\
bone & string & Bone hit (only if \textit{ot} true). \\
blind & bool & Indicates whether the shot was possible from the server’s perspective. A line is traced between the shooter position and the target. The value is true if the shot was not possible (i.e., a blind shot), and false otherwise (only if \textit{ot} true). \\
visible\_players & int32[] & List of the players in the shooter's field of vision at the time of the firing. Visible players are identified by their player\_id. \\[4pt]

% ======================= HealthUpdate =======================
\multicolumn{3}{l}{\textbf{Message: HealthUpdate}} \\
player\_id & int32 & Player ID. \\
old & int32 & Previous health value. \\
new\_ & int32 & Updated health value. \\[4pt]

% ======================= CheatEvent =======================
\multicolumn{3}{l}{\textbf{Message: CheatEvent}} \\
cheater\_id & int32 & Cheater player ID. \\
target\_id & int32 (opt) & Target player's ID (if the cheat is launched against a target). \\
type & string & Type of cheat. \\
start & bool & Whether cheat started/stopped. \\
dur & float (opt) & Duration of cheat. \\
out & int32 (opt) & Outgoing damage loss (For DDoS). \\
in\_ & int32 (opt) & Incoming damage loss (For DDoS). \\[4pt]

\bottomrule
\label{tab:gameevents}
\end{longtable}

\subsection{Packet Logs}\label{app:packetLogs}

\textit{opt} : optional

\begin{longtable}{p{3.5cm} p{3cm} p{8cm}}

\toprule
\textbf{Field} & \textbf{Type} & \textbf{Description} \\
\midrule
\endhead

% ======================= Packet =======================
\multicolumn{3}{l}{\textbf{Message: Packet}} \\
ts & Timestamp & Packet reception timestamp. \\
ind & int32 & Index in ring buffer. \\
frame & int32 & Frame counter. \\
incoming & bool & Outgoing (false), incoming (true). By default, we only add \texttt{Header}, \texttt{notification} and \texttt{Bunch}, for incoming packets. the content of outgoing packets can be retrieved as an incoming packet on the other endpoint, the hash can be leveraged to verify the packet's integrity. \\
header & Header (opt) & Packet header. \\
notification & Notification (opt) & Notification payload. \\
bunches & Bunch[] & Grouped payload structures. \\
hash\_data & bytes & hash of the packet's data, to compare packets at client/server and verify no tampering \\[4pt]
connection\_error & ConnectionError (opt) & Connection-level error info. \\[4pt]

% ======================= Header =======================
\multicolumn{3}{l}{\textbf{Message: Header}} \\
ids & HeaderIDs (opt) & Session and client IDs. \\
handshake & HeaderHandshake (opt) & Handshake metadata. \\[4pt]

% ======================= HeaderIDs =======================
\multicolumn{3}{l}{\textbf{Message: HeaderIDs}} \\
sessionid & uint32 & Server-assigned ID that increments on every non-seamless travel per connection, identifying the game session. Internally used to distinguish between old/new connection with the same IP:Port. \\
clientid & uint32 & Client-assigned connection ID that increments on each connection to a server. \textbf{Note:} This is different from \texttt{player\_id}; it is an internal Unreal Engine counter used to distinguish connections, not individual players. \\
address\_id & int32 & ID of the IP address of the corresponding player (the address used can be retrieved by mapping the id and new\_address events). \\
hash\_data & bytes & hash of the packet's data, to compare packets at client/server and verify no tampering \\[4pt]

% ======================= HeaderHandshake =======================
\multicolumn{3}{l}{\textbf{Message: HeaderHandshake}} \\
packet\_type & int32 & Type of handshake packet. \\
sent\_packet\_count & int32 & Number of packets sent during handshake. \\
restart & bool & Restart handshake flag. \\
secretid & int64 & Secret key used. \\
ts & Timestamp & Timestamp exchanged. \\
cookie & string & Challenge cookie. \\[4pt]

% ======================= Notification =======================
\multicolumn{3}{l}{\textbf{Message: Notification}} \\

type & string & Notification type. \\
seq & int32 & Sequence number, incremented every packet per flow. in the range [0, 65535]. \\
acked\_seq & int32 & Acknowledged sequence number (last seq received, per flow). \\
extra\_info\_header & ExtraInfoNotification- Header (opt) & Additonal information. \\
server\_frame\_time & ServerFrameTime (opt) & Server frame time. \\[4pt]

% ======================= ExtraInfoNotificationHeader =======================
\multicolumn{3}{l}{\textbf{Message: ExtraInfoNotificationHeader}} \\
type & string & Extra information type. \\
jitter\_clock & int32 & Jitter clock value transmitted. \\[4pt]

% ======================= ServerFrameTime =======================
\multicolumn{3}{l}{\textbf{Message: ServerFrameTime}} \\
type & string & Entry type. \\
srvft & int32 & Server frame time. \\[4pt]

% ======================= Bunch =======================
\multicolumn{3}{l}{\textbf{Message: Bunch}} \\
payload & \textit{oneof} \{ \\
bunch\_actor\_channel & BunchActorChannel & Bunch containing actor data. \\
control\_channel & ControlChannel & Control channel information. \\
movement\_srv\_rcvd & MovementSrvRcvd & Movement received at the server level. \\
movement\_inv\_timestamp & MovementInvTimestamp & Invalid timestamp was received. \\
received\_property & ReceivedProperty& Unreal Engine property received. \\
invalid\_rep\_field & InvalidRepField & Invalid replication field received. \\
received\_rpc & ReceivedRPC & Remote Procedure Call received.  \\
ability\_fail & AbilityFail & error message: an ability failed to invoke. \\
\} \\
[4pt]

% ======================= BunchActorChannel =======================
\multicolumn{3}{l}{\textbf{Message: BunchActorChannel}} \\
bopen & bool & Channel opened. \\
bclose & bool & Channel closed. \\
breliable & bool & Reliable channel. \\
bpartial & bool & Partial  (fragmentation). \\
bpartial\_initial & bool & Start of partial bunch. \\
bpartial\_final & bool & End of partial bunch. \\
actor\_name & string & Actor name. \\
index & int32 & Actor channel index. \\[4pt]

% ======================= ControlChannel =======================
\multicolumn{3}{l}{\textbf{Message: ControlChannel}} \\
type & string & Control message type. \\
messagetype & string & Content of the control message. \\[4pt]

% ======================= MovementSrvRcvd =======================
\multicolumn{3}{l}{\textbf{Message: MovementSrvRcvd}} \\
type & string & Movement update type. \\
ts\_client & float & Client timestamp (seconds since the start of the game). \\
accel & string & Acceleration. \\
velocity & string & Velocity. \\
loc & string & Location. \\
rot & string & Rotation. \\
dt & float & Delta time with previous movement. \\[4pt]

% ======================= MovementInvTimestamp =======================
\multicolumn{3}{l}{\textbf{Message: MovementInvTimestamp}} \\
type & string & Update type. \\
ts\_client & float & Client timestamp (seconds since the start of the game). \\
ts\_server & float & Server timestamp (seconds since the start of the game). \\[4pt]

% ======================= ReceivedProperty =======================
\multicolumn{3}{l}{\textbf{Message: ReceivedProperty}} \\
type & string & Property type. \\
obj & string & Target object. \\
func & string & Property function. \\[4pt]

% ======================= InvalidRepField =======================
\multicolumn{3}{l}{\textbf{Message: InvalidRepField}} \\
obj & string & Invalid replicated field. \\[4pt]

% ======================= ReceivedRPC =======================
\multicolumn{3}{l}{\textbf{Message: ReceivedRPC}} \\
type & string & RPC type. \\
func & string & Function name. \\
obj & string & Target object. \\
delayed & bool & Whether RPC was delayed. \\
parms\_nbr & int32 & Number of parameters. \\
parms\_size & int32 & Size of parameter data. \\
handle & int32 (opt) & Optional handle, only for specific RPC. \\
input & bool (opt) & Optional input flag, only for specific RP. \\[4pt]

% ======================= AbilityFail =======================
\multicolumn{3}{l}{\textbf{Message: AbilityFail}} \\
type & string & Event type. \\
ability & string & Name of ability. \\
reason & string & Reason for failure. \\[4pt]

% ======================= ConnectionError =======================
\multicolumn{3}{l}{\textbf{Message: ConnectionError}} \\
type & string & Error type. \\
err & int32 & Error code. \\
fatal & bool & Error is fatal (leads to a disconnection). \\

\bottomrule
\label{tab:PacketLogs}
\end{longtable}

\twocolumn
\section{Feedback form}\label{sec:feedback_form}

\textbf{Feedback Form}

Thank you for participating in our video game test! Your feedback is valuable for our research. Please answer the following questions based on your experience.

\subsection*{Consent}

This study collects your IP address for research purposes. Do you consent to this data being collected and used? 

\subsection*{Player Information}

\begin{enumerate}
    \item What country are you playing from?
    
    \item What is the Zip-Code of your location?
    
    \item How would you rate your skill level in shooter games? (1 = beginner, 2 = average, 3 = experienced)
    
    \item If you have a ranking in a popular game, please write it below (e.g., CS:GO rank) 
\end{enumerate}

\subsection*{Performance \& Stability}

\begin{enumerate}
    \setcounter{enumi}{4} % Continue numbering from 5
    \item Did you encounter any crashes? (Yes/No) 
    \begin{itemize}
        \item If yes, please describe when it happened (in the gameplay) and send us the game logs by zipping the log folder \textit{Linux/LyraStarterGame/Saved/}
    \end{itemize}

    \item Did you experience any lag / frame rate drops (e.g., delayed input / character teleporting)? (Yes/No) 
    \begin{itemize}
        \item If yes, how severe was it? Describe it briefly: 
    \end{itemize}

    \item How would you rate the overall visual performance? (Poor, Fair, Good, Excellent) 
    \begin{itemize}
        \item Did you encounter any graphical glitches or artifacts? (Yes/No) 
        \item If yes, please describe them: 
    \end{itemize}
\end{enumerate}

\subsection*{Additional Comments}

\begin{enumerate}
    \setcounter{enumi}{12} % Continue numbering from 12
    \item Is there anything else you'd like to share about the game's performance?
\end{enumerate}

\end{document}